\documentclass[a4paper,fleqn,usenatbib]{mnras}

\usepackage[T1]{fontenc}
\usepackage{ae,aecompl}

\usepackage{graphicx}	
\usepackage{amsmath}	
\usepackage{amssymb}	
\usepackage{color}

\definecolor{grey}{rgb}{0.7,0.7,0.7}

\newcommand{\bluetides}{{\tt BlueTides}}
\newcommand{\webb}{{\em Webb}}
\newcommand{\webbtelescope}{{\em Webb Telescope}}

\newcommand{\hubble}{{\em Hubble}}
\newcommand{\spitzer}{{\em Spitzer}}

\title[Nebular Line Emission During the EoR]{Nebular Line Emission During the Epoch of Reionization}

\author[Stephen M. Wilkins et al.]{
Stephen M. Wilkins,$^{1}$\thanks{E-mail: s.wilkins@sussex.ac.uk}
Christopher C. Lovell,$^{1}$ 
Ciaran Fairhurst,$^{1}$ 
Yu Feng,$^{2}$ \newauthor 
Tiziana Di Matteo,$^{3,4}$ 
Rupert Croft,$^{3,4}$  
Jussi Kuusisto,$^{1}$ 
Aswin P. Vijayan,$^{1}$ \newauthor 
Peter Thomas,$^{1}$
\\
$^1$\,Astronomy Centre, Department of Physics and Astronomy, University of Sussex, Brighton, BN1 9QH, UK \\
$^3$\,Berkeley Center for Cosmological Physics, University of California, Berkeley, Berkeley CA, 94720, USA \\
$^2$\,McWilliams Center for Cosmology, Carnegie Mellon University, Pittsburgh PA, 15213, USA \\
$^4$\,School of Physics, University of Melbourne, VIC 3010, Australia.\\
}

\date{Accepted XXX. Received YYY; in original form ZZZ}

\pubyear{2019}

\begin{document}
\label{firstpage}
\pagerange{\pageref{firstpage}--\pageref{lastpage}}
\maketitle


\begin{abstract}
Nebular emission lines associated with galactic H{\sc ii} regions carry information about both physical properties of the ionised gas and the source of ionising photons as well as providing the opportunity of measuring accurate redshifts and thus distances once a cosmological model is assumed. While nebular line emission has been extensively studied at lower redshift there are currently only few constraints within the epoch of reionisation (EoR, $z>6$), chiefly due to the lack of sensitive near-IR spectrographs. However, this will soon change with the arrival of the \webbtelescope\ providing sensitive near-IR spectroscopy covering the rest-frame UV and optical emission of galaxies in the EoR. In anticipation of \webb\ we combine the large cosmological hydrodynamical simulation \bluetides\ with photoionisation modelling to predict the nebular emission line properties of galaxies at $z=8\to 13$. We find good agreement with the, albeit limited, existing direct and indirect observational constraints on equivalent widths though poorer agreement with luminosity function constraints.
\end{abstract}

\begin{keywords}
galaxies: high-redshift -- galaxies: photometry -- methods: numerical -- galaxies: luminosity function, mass function
\end{keywords}

\section{Introduction}

Massive stars and active galactic nuclei (AGN) are often intense sources of Lyman-continuum (LyC, or hydrogen ionising) photons resulting in the formation of regions of ionised gas in their surroundings (e.g. H{\sc ii} regions). The emission from these regions carries information about the physical conditions in the interstellar medium (ISM) as well as the source of the ionising photons themselves. Key properties that can be constrained include the star formation rate \citep[e.g.][]{KE12, WLS2019}, gas metallicity \citep[e.g.][]{Tremonti2004}, temperature, density, dust content \citep[e.g.][]{Reddy2015}, and the presence of an AGN  \citep[e.g.][]{Baldwin1981}. Nebular line emission also enables the accurate measurement of redshifts, and thus distances once a cosmological model is assumed.

While there has been extensive progress in observing line emission at low \citep[e.g.][]{Brinchmann2004} and intermediate \citep[e.g.][]{Steidel1996, Shapley2003} redshifts there are few direct constraints at high-redshift. This is predominantly due to lack of sensitive near-IR spectrographs, particularly at $>2\mu$m where the rest-frame optical lines lie at $z>4$, and the comparative lack of strong lines, other than Lyman-$\alpha$, in the rest-frame UV. The small number of detections at high-redshift come overwhelmingly from Lyman-$\alpha$ \citep[e.g.][]{Stark2010, Pentericci2011, Stark2011, Caruana2012, Stark2013, Finkelstein2013, Caruana2014, Stark2017} though there has now been a handful of detections of the [C{\sc iv}]$\lambda 1548$ and [C{\sc iii}],C{\sc iii}]$\lambda 1907,1090$ lines \citep{Stark2015a, Stark2015b, Stark2017}. The presence of extremely strong optical lines can also be indirectly inferred from their impact on broadband photometry \citep[e.g.][]{Schaerer2010, Stark2013, Wilkins2013d, Smit2014, Wilkins2016c, deBarros2019} yielding constraints now available up to $z\approx 8$ \citep{deBarros2019}.

While existing observational constraints in the EoR are limited this will soon change with the arrival of the \webbtelescope. \webb's NIRSpec instrument will provide deep near-IR single slit, multi-object, and integral field spectroscopy from $\sim 0.7-5\mu$m, while the NIRISS and NIRCam instruments will, together, provide wide field slitless spectroscopy over a similar range. This is sufficient to encompass all the strong optical lines to $z\sim 7$ with [O{\sc ii}] potentially accessible to $z\sim 12$. \webb's mid-infrared instrument (MIRI) will provide mid-IR single slit, and slitless spectroscopy at $\lambda>5\mu$m, albeit at much lower sensitivity and thus will likely only detect line emission for the brightest sources in the EoR.

The existing direct and indirect constraints and the nearing prospect of \webb\ motivates us to produce predictions for the nebular emission line properties of galaxies in the EoR. In this paper, we combine the large \bluetides\ hydrodynamical simulation with photoionisation modelling to predict the nebular line properties of galaxies in the EoR, specifically ($z=8\to 13$). As part of this paper we also explore some of the photon-ionisation modelling assumptions including the choice of stellar population synthesis (SPS) model and initial mass function (IMF). These predictions can be used to optimise the design of \webb\ surveys prior to launch, targeting emission lines in the EoR. The observation of these lines will also provide a powerful constraint on the physics incorporated into galaxy formation models.

This work builds upon other recent efforts to model nebular lines using both simple analytical models \citep[e.g.][]{CL2001,Schaerer2009,Inoue2014,Gutkin2016,Feltre2016,Nakajima2018} and the modelling in semi-analytical galaxy formation models \citep[e.g.][]{Orsi2014} and hydrodynamical simulations \citep[e.g.][]{Shimizu2016, Hirschmann2017,Hirschmann2019}.

This article is structured as follows: in Section \ref{sec:BT} we describe the \bluetides\ simulation and our methodology for calculating spectral energy distribution including nebular emission (\S\ref{sec:BT.SED}). In Section \ref{sec:predictions} we present our predictions. In this section we also explore the impact of some modelling assumptions (\S\ref{sec:predictions.modelling}), including the choice of stellar population synthesis (SPS) model (\S\ref{sec:predictions.modelling.SPS}), initial mass function (IMF, \S\ref{sec:predictions.modelling.IMF}), photoionisation modelling parameters, including the impact of dust (\S\ref{sec:predictions.modelling.photo}). In Section \ref{sec:predictions.obs} we compare our predictions to existing observational constraints. In Section \ref{sec:c} we present our conclusions.


\section{Modelling Nebular Emission in BLUETIDES}\label{sec:BT}

\subsection{The BlueTides Simulation}

The \bluetides\ simulation\footnote{\url{http://bluetides-project.org/}} \citep[see][for description of the simulation physics]{Feng2015,Feng2016} is an extremely large cosmological hydrodynamical simulation designed to study the early phase of galaxy formation and evolution with a specific focus on the formation of the massive galaxies. \bluetides\ phase 1 evolved a $(400/h\approx 577)^{3}\,{\rm cMpc^3}$ cube to $z=8$ using $2\,\times\, 7040^{3}$ particles assuming the cosmological parameters from the {\em Wilkinson Microwave Anisotropy Probe} ninth year data release \citep{Hinshaw2013}. The dark matter particle and gas particle initial masses are $1.2\times 10^{7}h^{-1}\ {\rm M_{\odot}}$ and $2.36\times 10^{6}h^{-1}\ {\rm M_{\odot}}$ respectively. The gravitational softening length is $\epsilon_{\rm grav}=1.5 h^{-1}\ {\rm kpc}$. This is sufficient to allow us to easily resolve galaxies to $M_{\rm halo}\approx 10^{9}\,{\rm M_{\odot}}$ though in this work we adopt a more conservative approach only focussing on galaxies with stellar masses $>10^8\,{\rm M_{\odot}}$ which contain at least approximately 100 star particles in order to star formation history samplings effects (see Appendix \ref{sec:appendix.sampling} for exploration of sampling effects). The properties of galaxies in the simulation are extensively described in \citet{Feng2015,Feng2016, Wilkins2016b, Wilkins2016c, Waters2016a, Waters2016b, DiMatteo2016, Wilkins2017a, Wilkins2018}. While \bluetides\ contains super-massive black holes, and even a small number of AGN dominated sources at $z=8$, in this work we focus on the line emission arising solely from gas excited by stellar sources.

\subsubsection{Ages and Metallicities of Galaxies in BLUETIDES}\label{sec:BT.age_Z}

\begin{figure}
\centering
\includegraphics[width=20pc]{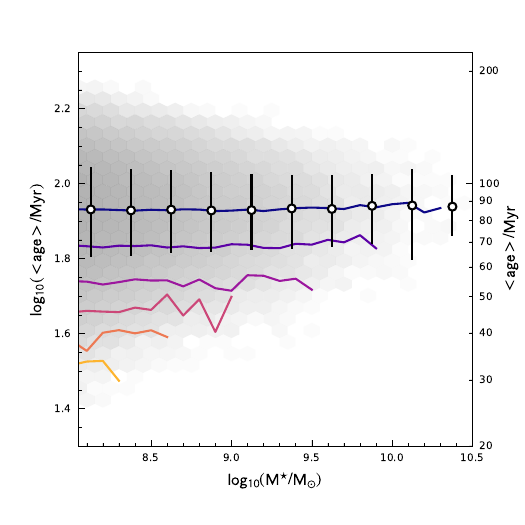}
\includegraphics[width=20pc]{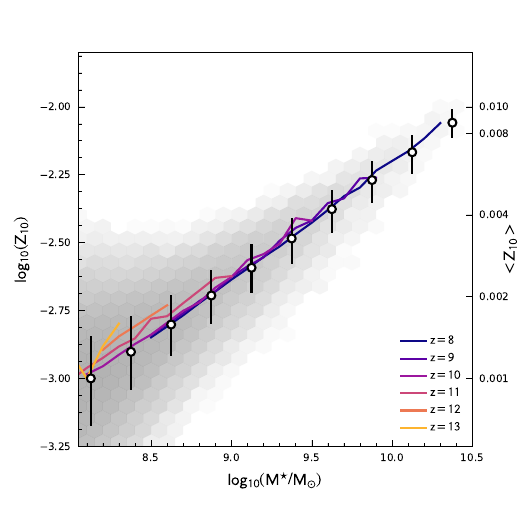}
\caption{The mass-weighted average age (top) and average metallicity (expressed as the mass fraction in elements heavier than helium) of star particles with ages $<10\,{\rm Myr}$ (bottom) as a function of stellar mass for $z\in\{8,9,10,11,12,13\}$. The solid lines show the median age/metallicity in $0.1\,{\rm dex}$ wide $\log_{10}(M_{\star}/{\rm M_{\odot}})$ bins. The 2D histogram shows the distribution of stellar masses and mean ages/metallicities at $z=8$ using a logarithmic scale.}
\label{fig:ages_metallicities}
\end{figure}

As emission line luminosities and equivalent widths are predominantly driven by galaxy star formation and metal enrichment histories it is useful to explore the average ages and metallicities predicted by \bluetides. The mean stellar age and mean metallicity of young ($<10\,{\rm Myr}$) stars are shown as a function of stellar mass for a range of redshifts in Figure \ref{fig:ages_metallicities}. These correlations were previously discussed in more detail in \citet{Wilkins2017a} while a more detailed analysis of the joint star formation and metal enrichment history is presented in Fairhurst et al.\ {\em in-prep}. In short, the mean stellar age appears to show little dependence on mass but evolves strongly with redshift while the mean metallicity of young stars shows a power-law dependence ($Z\propto M^{0.4}$) on stellar mass but little evolution with redshift.

\subsection{Spectral Energy Distribution Modelling}\label{sec:BT.SED}

We model the spectral energy distributions (SEDs) of galaxies in \bluetides\ as the sum of the SEDs of each star particle identified as belonging to each galaxy.

We begin by associating each star particle with a {\em pure} stellar SED according to its age and metallicity. To obtain this SED we interpolate publicly available grids produced by stellar population synthesis (SPS) models. By default we make the following modelling choices: we assume the {\sc bpass} v2.2.1 SPS model \citep{SE2018, Eldridge2017}\footnote{\url{https://bpass.auckland.ac.nz}} and a modified version of the Salpeter IMF containing a flattened ($\alpha=-1.3$) power-law at low-masses ($m<0.5\,{\rm M_{\odot}}$). This IMF produces very similar ($<0.05\,{\rm dex}$) results to the assumption of a \citet{Chabrier2003} IMF but permits a fairer comparison with the alternative IMFs available for \textsc{bpass}. In \S\ref{sec:predictions.modelling} we explore the impact of these, and other assumptions.

\subsubsection{Nebular emission}

Using the intrinsic stellar SED, and assuming no escape of LyC photons, we use the \textsc{cloudy} photoionisation code \citep{Cloudy17}\footnote{\url{https://www.nublado.org}} to associate each young ($t<10\,{\rm Myr}$) star particle with an individual H\textsc{ii} region or birth cloud. The metallicity of this region is assumed to be identical to the star particle itself and we adopt the same interstellar abundances and dust depletion factors as described in \citet{Gutkin2016}.

To model the nebular emission associated with a stellar population we adopt a similar approach to \citet{CL2001} \citep[see also][]{Gutkin2016, Feltre2016}. Like these works we choose characterise our photoionisation modelling using the density of hydrogen ($n_H$) and ionisation parameter at the Stromgren radius $U_S$. This is defined as,
\begin{equation}
U_{S} \propto\left(Q\epsilon^2 n_H\right)^{1/3}
\end{equation}
where $\epsilon$ is the effective gas filling factor.

We differ from previous approaches by parameterising models for the ionising spectrum in terms of an ionisation parameter defined at a reference age ($t=1\,{\rm Myr}$) and metallicity ($Z=0.02$) - $U_{S, {\rm ref}}$. Because of this the actual ionisation parameter passed to \textsc{cloudy} depends on the ionising photon production rate relative to the reference value, i.e.
\begin{equation}
U_{S} = U_{S, {\rm ref}}\left(\frac{Q}{Q_{\rm ref}}\right)^{1/3}.
\end{equation}
This ensures that the assumed geometry of the H{\sc ii} region, encoded in the $\epsilon^2 n_H$ term, is fixed for different metallicities/ages. By default we assume $\log_{10}(U_{S,{\rm ref}})=-2$ and $\log_{10}(n_{H}/{\rm cm^{-3}})=2.5$.

In our modelling we include the effect of dust grains which can not only boost certain lines \citet[see][]{vanHoof2004, Nakajima2018} but also provide an additional source of attenuation. Specifically we include \textsc{cloudy}'s default implementation of Orion-type graphite and silicate grains but scale the abundances to match those assumed for carbon and silicon in the H\textsc{ii} region. The impact of the inclusion of grains is discussed in more detail in \S\ref{sec:predictions.modelling.photo}.

In our calculations we assume the default {\sc cloudy} stopping temperature (4000K) which is suitable for UV/optical recombination lines.

\subsubsection{Modelling attenuation by dust in the ISM}

\bluetides\ includes a simple model to account for dust in the wider intervening ISM. This ISM dust component is modelled using a simple scheme which links the smoothed metal density integrated along a single consistent line-of-sight to each star particle within each galaxy to the dust optical depth in the $V$-band ($550\,{\rm nm}$). Attenuation at other wavelengths is determined using a simple attenuation curve of the of the form,
\begin{equation}
\tau_{\lambda} = \tau_V\times (\lambda/550{\rm nm})^{-1}.
\end{equation}
This model has a single free parameter which effectively links the surface density of metals to the optical depth. This parameter is tuned to recover the shape of the of the observed $z=8$ far-UV luminosity function. For a full description see \citet{Wilkins2017a}.

\section{Predictions for BLUETIDES}\label{sec:predictions}

\begin{figure}
\centering
\includegraphics[width=20pc]{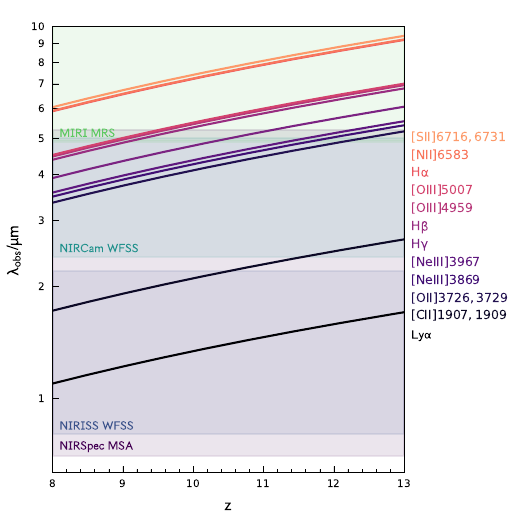}
\caption{The observed wavelength of the twelve lines considered in this work along with the ranges probed by various \webb\ instruments.}
\label{fig:line_visibility}
\end{figure}

Using the methodology outlined above we calculate the luminosities and equivalent widths of twelve prominent rest-frame UV and optical single lines or doublets (see Fig. \ref{fig:line_visibility} for a list of lines and their accessibility at high-redshift to \webb) for all galaxies in \bluetides\ at $z=8-13$ with $M_{\star}>10^{8}\, {\rm M_{\odot}}$. Equivalent widths (EWs) are calculated simply by dividing the line luminosities by the underlying continuum emission (which includes contributions from both transmitted starlight and nebular continuum emission).

\subsection{Line Luminosities and Equivalent Width Distributions}

\begin{figure}
\centering
\includegraphics[width=20pc]{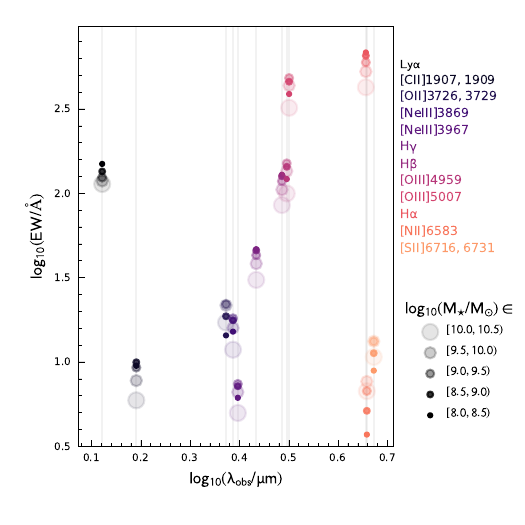}
\caption{Predicted equivalent widths at $z=8$ for various stellar mass bins of the 12 lines or line doublets considered in this work.}
\label{fig:predictions_summary}
\end{figure}

Detailed diagnostic plots for each of the 12 calculated single lines or doublets are presented in Appendix \ref{sec:detailed_predictions}. Tabulated results for all 12 lines are also all available in electronic form at \url{https://github.com/stephenmwilkins/BluetidesEmissionLines_Public}. Fig. \ref{fig:predictions_summary} provides a summary showing the median equivalent widths of all 12 lines at $z=8$ in bins of stellar mass.

\begin{figure*}
\centering
\includegraphics[width=40pc]{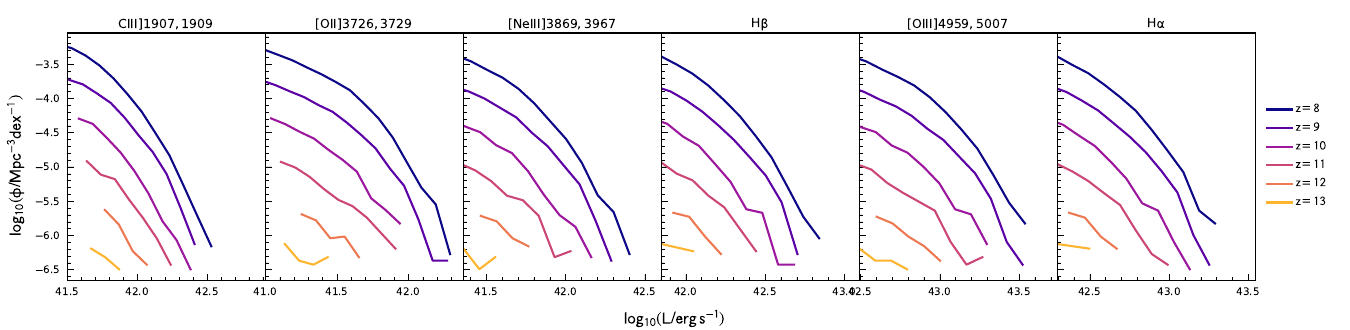}
\includegraphics[width=40pc]{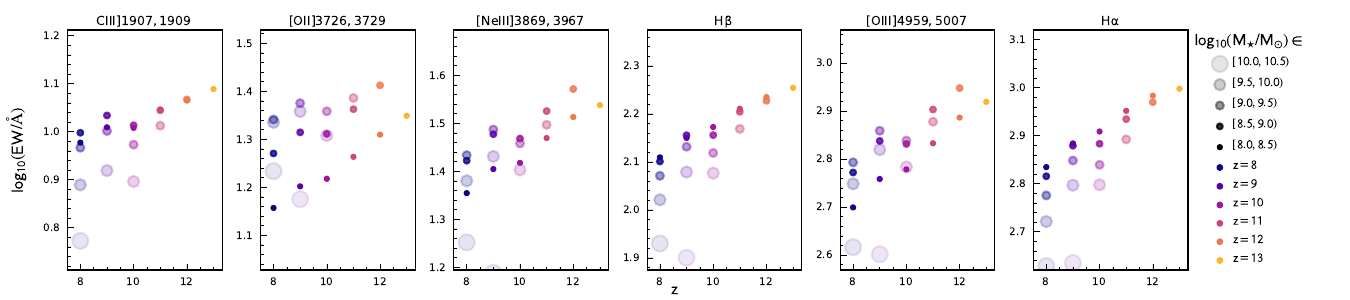}
\includegraphics[width=40pc]{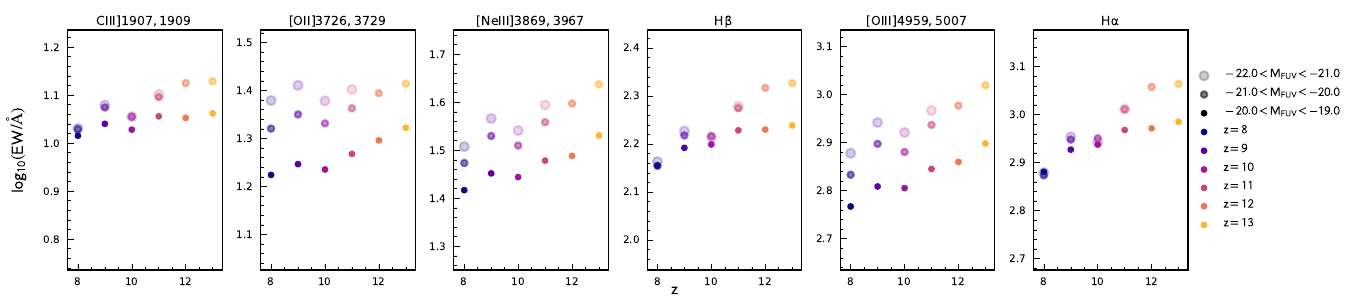}
\includegraphics[width=40pc]{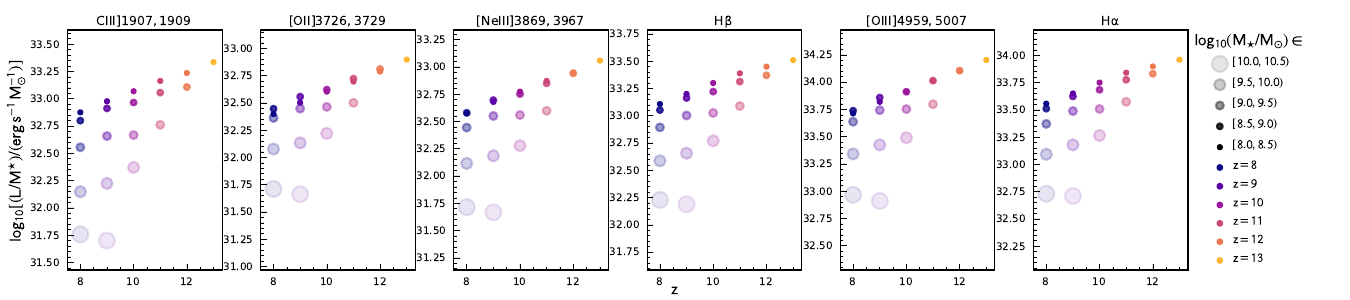}
\includegraphics[width=40pc]{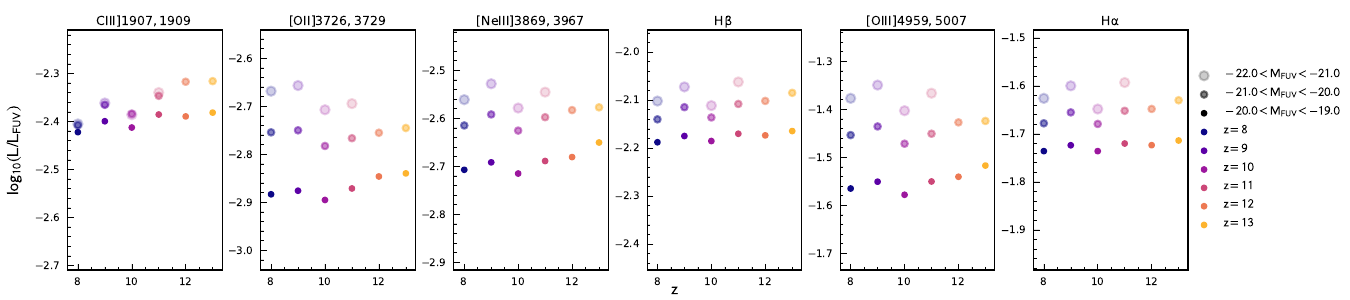}
\caption{Predictions for the properties of 6 prominent UV and optical lines in \bluetides\. In the top panel we show both the intrinsic and dust-attenuated luminosity functions for each line at $z\in\{8,9,10,11,12,13\}$. In the next two rows we show the median attenuated equivalent width in bins of stellar mass and UV luminosity respectively. In the fourth row we show the median specific line luminosity ($L/M_{\star}$) in stellar mass bins while in the final row we show the median ratio of the line luminosity to the UV luminosity in bins of UV luminosity.}
\label{fig:predictions}
\end{figure*}

A more detailed summary, concentrating on only 6 lines or doublets\footnote{Here we have combined both the [Ne{\sc iii}] and [O{\sc iii}] doublets into a single quantity.}, is presented in Fig. \ref{fig:predictions}. Here we show predictions for both the luminosity function, and EW, $L/M_{\star}$, and $L/L_{\rm FUV}$ distributions as a function of redshift, stellar mass, and FUV luminosity.

The luminosity function (Fig. \ref{fig:predictions}, row 1) of each line broadly follows a similar trend to the UV luminosity function: intrinsically the LF is approximated by a single power-law; the inclusion of dust however causes a strong break at high luminosities. Like the UV LF the line luminosity function evolves strongly with redshift, increasing by a factor $\approx 1000$ from $z=13\to 8$. At fixed stellar mass equivalent widths (Fig. \ref{fig:predictions}, row 2) mostly increase to higher-redshift. This predominantly reflects that higher-redshift galaxies are generally younger (see \S\ref{sec:BT.age_Z}) and thus their SED has a larger contribution from the most massive (LyC producing) stars. This also results in a wavelength dependence with the EWs of bluer lines evolving less strongly with redshift. The trend of line EW with stellar mass is more complex due to the correlation of stellar mass with metallicity (\S\ref{sec:BT.age_Z}). For example, for this reason the EW of the hydrogen recombination lines drops at higher stellar mass while that of the [O{\sc ii}]$3726,3729$ line peaks at $M^{\star}\sim 10^{9.5}\,{\rm M_{\odot}}$ (at $z=8$). The trend of EW with UV luminosity (Fig. \ref{fig:predictions}, row 3) shows a similar trend with redshift. However, the trend with UV luminosity is less pronounced compared to with stellar mass due to the weaker correlation between observed UV luminosity and metallicity. The specific line luminosity ($L/M_{\star}$) (Fig. \ref{fig:predictions}, row 4) shows a clear increase to higher-redshift, again this is driven by the fact that at higher-redshift the average age of the stellar populations are typically younger and thus produce more ionising photons per unit stellar mass. There is also a strong trend with stellar mass. While some of this is affected by metallicity it is predominantly dominated by the effect of dust. In contrast to the other quantities the ratio of line luminosity to FUV luminosity (Fig. \ref{fig:predictions}, row 5) shows little evolution with redshift. This is because both ionising and FUV photons are dominated by the most massive stars. There is however a strong FUV luminosity dependence due to increased effect of dust.

\subsection{Impact of Modelling Assumptions}\label{sec:predictions.modelling}

These predictions depend not only simulation physics but the additional modelling assumptions made in \S\ref{sec:BT.SED}. In this section we explore the impact of some of these assumptions.

\subsubsection{Stellar Population Synthesis Model}\label{sec:predictions.modelling.SPS}

\begin{figure*}
\centering
\includegraphics[width=40pc]{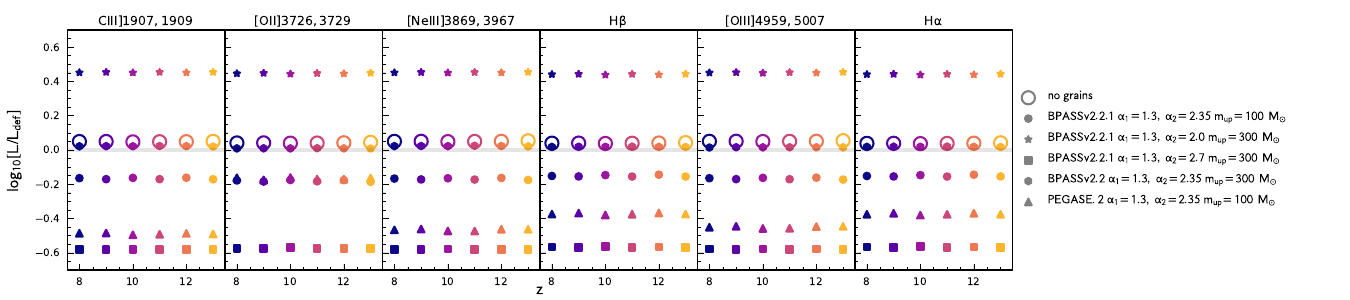}
\includegraphics[width=40pc]{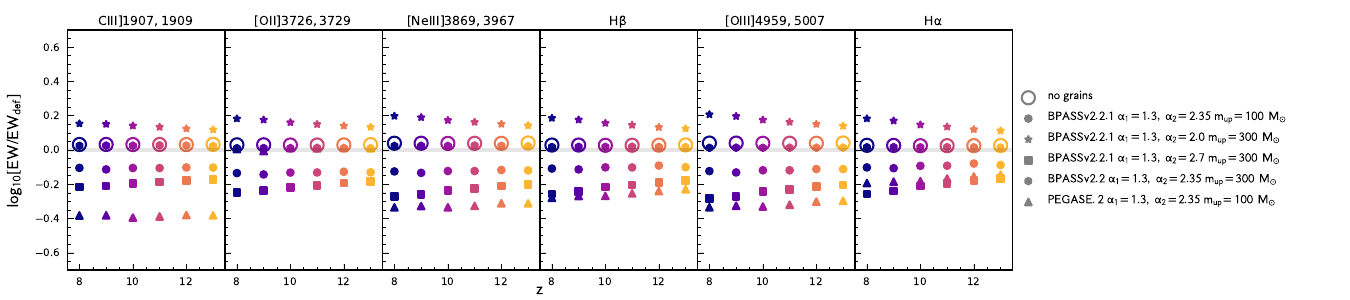}
\caption{The impact of changing various assumptions (SPS model, IMF, inclusion of grains) on the average luminosities (top panels) and equivalent widths (bottom panels) of the 6 lines relative to our default choices (BPASSv2.2.1, $\alpha_2=2.35$, $m_{\rm up}=300\,{\rm M_{\odot}}$).}
\label{fig:assumptions_impact}
\end{figure*}

The production rate of LyC photons and the shape of the ionising spectrum (or hardness) predicted for a given simple stellar population is sensitive to the a range of stellar evolution and atmosphere modelling assumptions and thus choice of stellar population synthesis (SPS) model (see \S\ref{sec:photo.ion.SPS} for more details). Fig. \ref{fig:assumptions_impact} shows the impact of changing the assumed SPS model on the line luminosities and equivalent widths predicted by \bluetides. Adopting the previous release (2.2) of {\sc bpass} produces only relatively small changes ($<0.1\,{\rm dex}$) to the predicted line luminosities and equivalent widths. On the other hand adopting the {\sc Pegase}.2 SPS model (and $m_{\rm up}=100\,{\rm M_{\odot}}$) produces a significant decrease in the luminosities and equivalent widths relative to our default model. For most lines luminosities drop by $\sim 0.5$ dex while equivalent widths drop by $\sim 0.3$. While some of this decrease can be attributed to the small upper-mass cutoff of the IMF most of the effect is attributed to wider modelling differences between {\sc Pegase}.2 and {\sc bpass}, in particular the inclusion of binary stars in the latter.

\subsubsection{Initial Mass Function}\label{sec:predictions.modelling.IMF}

The production rate and hardness of LyC photons are also affected by the choice of initial mass function (see \S\ref{sec:photo.ion.IMF} for more details). Fig. \ref{fig:assumptions_impact} shows the impact of assuming an alternative IMF on the average predicted line luminosities and equivalent widths as a function of redshift. Unsurprisingly, increasing the fraction of high-mass stars, either by extending the high-mass cutoff or flattening the high-mass slope results in increased line luminosities and equivalent widths. As continuum luminosities are also increased the impact on equivalent widths is smaller than on line luminosities themselves. It is also important to note that flattening the IMF in this way would also increase the far-UV continuum luminosities of galaxies breaking the otherwise good agreement with observations \citep[see][]{Wilkins2017a}. This could however be ameliorated by having more aggressive dust attenuation. The effect of steepening the slope produces the opposite effect. Steepening the IMF to this extent will also significantly decrease the far-UV continuum luminosities again breaking the good agreement with observational constraints. In this case the good agreement can not be recovered without more drastic changes to the simulation physics.

\subsubsection{Photoionisation Modelling}\label{sec:predictions.modelling.photo}

The luminosity of each line is also sensitive the parameters encapsulating the geometry, density, excitation, and dust content/composition of the H{\sc ii} region. The impact of the ionisation parameter, which effectively encodes the geometry of the region, and the hydrogen density are discussed in \S\ref{sec:photo.photo.modelassumptions}. As demonstrated in Fig. \ref{fig:photo_params} the choice of these parameters can have a significant ($>\pm 0.5\ {\rm dex}$) impact on the luminosities and equivalent widths of metal lines.

As noted previously by default we include dust-grain physics. Within our model framework Fig. \ref{fig:const_dust_comparison} shows the impact of grains on the emergent line luminosities as a function of metallicity for a constant burst of star formation while in Fig. \ref{fig:assumptions_impact} we show the effect of turning off grain physics on our overall results. Removing grains results in a boost of $\sim 0.05$ dex to both luminosities and equivalent widths.

\subsection{Comparison to other models}\label{sec:predictions.models}

Like this work, \citet{Shimizu2016} (S16) model the UV/optical line emission of galaxies in the EoR by combining photonionsation modelling with a cosmological hydrodynamical simulation albeit with several key differences, including the base simulation physics, choice of SPS model and IMF, photoionisation model, and wider dust model. Overall we find good agreement between our predictions and S16.

\subsection{Comparison with existing observational constraints}\label{sec:predictions.obs}

As noted in the introduction there remain relativel few constraints (direct or otherwise) on optical/UV line emission at very high-redshift.

The majority of direct constraints come from observations of Lyman-$\alpha$ \citep[e.g.][]{Stark2010, Pentericci2011, Stark2011, Caruana2012, Stark2013, Finkelstein2013, Caruana2014, Stark2017}. However, the Lyman-$\alpha$ line is resonantly scattered by the ISM/IGM significantly complicating its modelling \citep[see][]{Smith2019}. For this reason we have omitted a detailed comparison with Lyman-$\alpha$ observations. We do however nevertheless make predictions for the Lyman-$\alpha$ properties including dust attenuation but not scattering by the ISM/IGM. These predictions are presented in Appendix \ref{sec:detailed_predictions}.

\begin{figure}
\centering
\includegraphics[width=20pc]{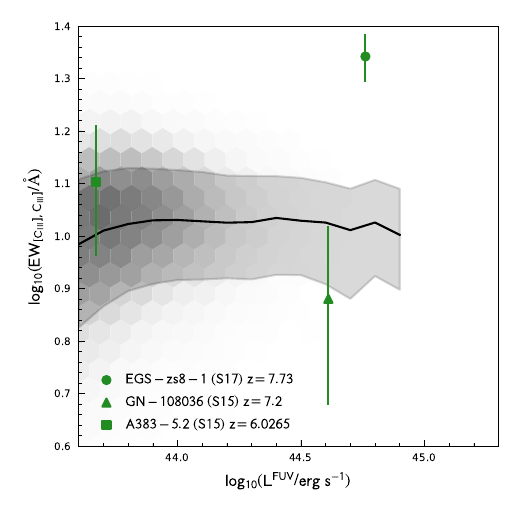}
\caption{The predicted distribution of [C{\sc iii}],C{\sc iii}] equivalent width and observed far-UV luminosity at $z\sim 8$. Points denote individual objects from $z=5-8$.}
\label{fig:CIIIobs}
\end{figure}

Recently \citet{Stark2015a} and \citet{Stark2017} have obtained constraints on the [C{\sc iii}],C{\sc iii}] doublet at $z=6-8$. These constraints are shown in Fig. \ref{fig:CIIIobs} alongside predictions from \bluetides. The two faintest objects (A383-5.2, GN-108036) have EWs statistically consistent with the \bluetides\ predictions assuming our default modelling choices. However, were we to alternatively assume the \textsc{Pegase.2} SPS model (see Fig. \ref{fig:assumptions_impact}) our predictions would lie below both these observations, albeit without strong statistical significance given the small number of objects and large measurement uncertainties. The brightest object \citep[EGS-zs8-1][]{Stark2017} lies $\sim 0.35$ dex above the median prediction for the same FUV luminosity. However, as this object is very bright this raises the possibility of a contribution from an AGN, which would raise the EW \citep[see e.g.][]{Nakajima2018}. While \bluetides\ includes AGN their contribution is not included in this work but instead deferred to a future study.

\begin{figure}
\centering
\includegraphics[width=20pc]{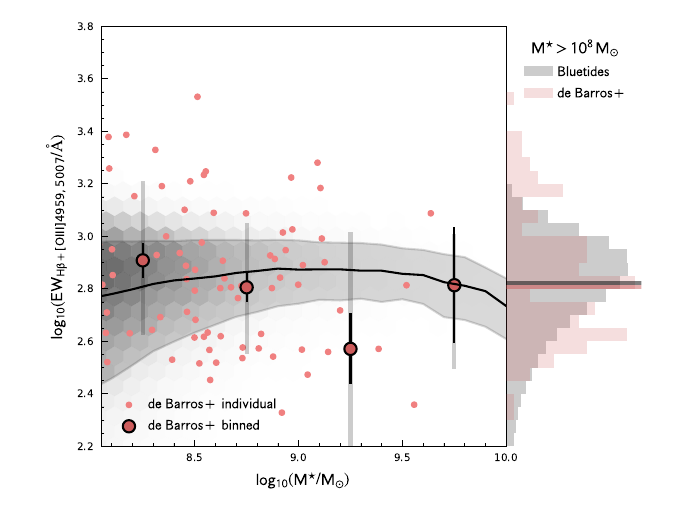}
\caption{The observed \citet{deBarros2019} and predicted distribution of combined H$\beta$ and [O{\sc iii}]$\lambda$4959,5007 equivalent widths and stellar masses at $z\sim 8$. The small red circles show the individual measurements from \citet{deBarros2019} while the large point denote the median value in 0.5 dex wide bins of stellar mass. The small and large error bars denote the error on the median and the 16-84th percentile range respectively. The dark and light solid lines show the intrinsic and attenuated predictions from \bluetides\ respectively. The histograms on the right hand side show the distribution of equivalent widths for galaxies with $M^{\star}>10^{8}\,{\rm M_{\odot}}$.}
\label{fig:EWdeBarros}
\end{figure}

Indirect constraints on the strength of the strongest optical lines are possible through the effect of these lines on broad-band photometry \citep[e.g.][]{Schaerer2010, Stark2013, Wilkins2013d, Smit2014, Wilkins2016c, deBarros2019}. \citet{deBarros2019} recently combined  \hubble\ and \spitzer\ observations probing the rest-frame UV and optical to constrain the prominent H$\beta$ and [O{\sc iii}]$\lambda$4959,5007 lines at $z\approx 8$. As shown in Fig. \ref{fig:EWdeBarros} the H$\beta$ $+$ [O{\sc iii}] EW distribution measured by \citet{deBarros2019} has an almost identical median to that predicted by \bluetides\ for our default assumptions. However we do fail to explain the tail of very-high ($>2000{\rm\AA}$) and low EW sources. As noted in \ref{sec:photo.photo.modelassumptions} the [O{\sc iii}]$\lambda$4959,5007 lines are sensitive to the choice of ionisation parameter. If we instead of a single reference ionisation parameter we chose a distribution this would naturally produce extended tails. It is also worth noting that if instead we adopted the \textsc{Pegase.2} SPS model our predictions would fall below these observational constraints (though this could be ameliorated by assuming a more high-mass biased IMF). Similarly, adopting a model in which line emission is more strongly attenuated by dust would only be consistent by also changing the IMF.

\begin{figure}
\centering
\includegraphics[width=20pc]{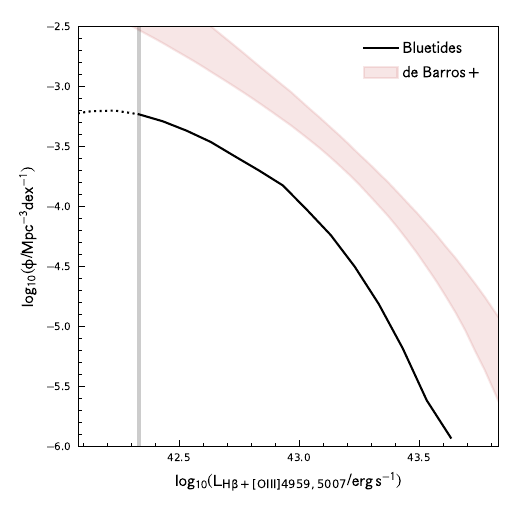}
\caption{The observed \citet{deBarros2019} and predicted combined H$\beta$ and [O{\sc iii}]$\lambda$4959,5007 line luminosity function at $z\sim 8$. The vertical line denotes the approximate completeness limit of the predicted line luminosity function.}
\label{fig:LFdeBarros}
\end{figure}

\begin{figure}
\centering
\includegraphics[width=20pc]{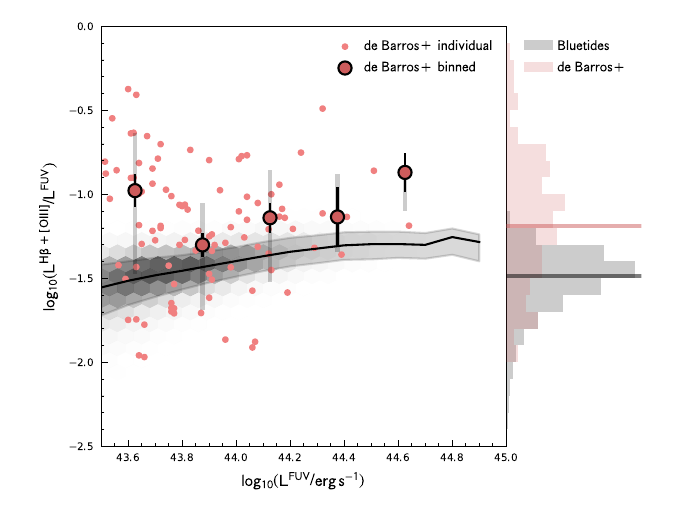}
\caption{The observed \citet{deBarros2019} and predicted distribution of the ratio of the H$\beta$ and [O{\sc iii}]$\lambda$4959,5007 line luminosities to the far-UV luminosity and far-UV luminosities at $z\sim 8$. The small red circles show the individual measurements from \citet{deBarros2019}. The dark and light solid lines show the intrinsic and attenuated predictions from \bluetides\ respectively. The histograms on the right hand side show the distribution of ratios for galaxies with $M^{\star}>10^{8}\,{\rm M_{\odot}}$.}
\label{fig:LUVLlineRatio}
\end{figure}

\begin{figure}
\centering
\includegraphics[width=20pc]{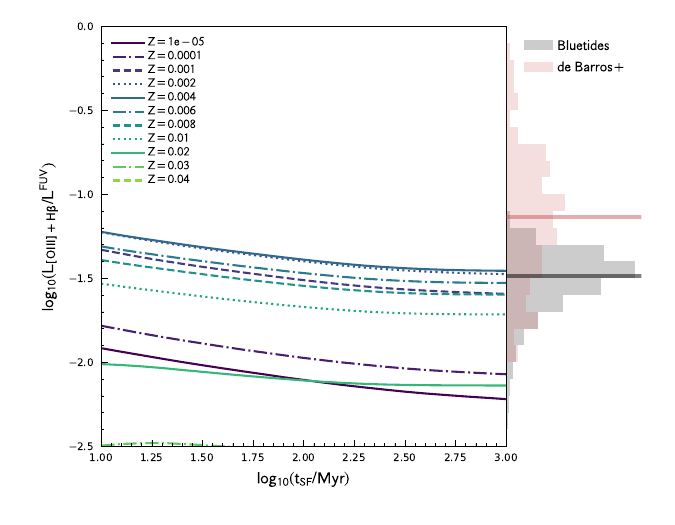}
\caption{The H$\beta$ and [O{\sc iii}]$\lambda$4959,5007 line luminosities - far-UV luminosity ratio measured by \citet{deBarros2019} compared to predictions from photoionisation modelling assuming a constant star formation activity ($t=10\to 1000\,{\rm Myr}$) and a range of metallicities. The histograms on the right hand side show both the observed and predicted distribution of ratios for galaxies with $M^{\star}>10^{8}\,{\rm M_{\odot}}$.}
\label{fig:LUVLlineRatioPhysics}
\end{figure}

Unfortunately this good agreement is not seen in the luminosity function, as shown in Fig. \ref{fig:LFdeBarros}. The observed luminosity function is systematically offset to higher luminosities ($\sim 0.4\,{\rm dex}$) or higher space densities ($\sim 0.7\,{\rm dex}$) than that predicted by \bluetides. The cause of this discrepancy appears to lie in the relation between the combined line luminosity and the far-UV luminosity, which is used by \citet{deBarros2019} to convert the observed FUV luminosity function to a line luminosity function. The individual values measured by \citet{deBarros2019} for this are shown in Fig. \ref{fig:LUVLlineRatio} and are compared to the values predicted by \bluetides. The measured values are on average $\sim 0.4\,{\rm dex}$ higher than predicted by \bluetides. As many of the measured values are above the intrinsic expectation (see Fig. \ref{fig:LUVLlineRatioPhysics}) one interpretation of this discrepancy is that \citet{deBarros2019} measure higher dust attenuations than predicted by \bluetides. It is also possible that differences between the measured and predicted values of the metallicity, age, ionisation parameter, and hydrogen density can have an effect. Given the limited observational constraints such differences may not be surprising considering the range of degeneracies present. As direct emission line measurements become available from the {\em Webb Telescope} and other upcoming facilities the cause of this discrepancy should become clearer.

\section{Conclusions}\label{sec:c}

Using the large cosmological hydrodynamical simulation \bluetides\ combined with photoionisation modelling we have made detailed predictions for the luminosities (including luminosity function) and equivalent widths of twelve prominent rest-frame UV and optical emission lines for galaxies with $M^{\star}>10^{8}\,{\rm M_{\odot}}$ across the EoR ($8<z<13$). As part of this analysis we also explored the impact of various modelling assumptions including the choice of stellar population synthesis model, initial mass function, and photoionisation modelling, finding that these can have a significant impact.

At present there are few observational constraints on line emission available in the EoR with only a handful of direct constraints on the [C{\sc iii}],C{\sc iii}] doublet along with indirect constraints on H$\beta$ and [O{\sc iii}]$\lambda$4959,5007 based on \spitzer\ photometry \citep{deBarros2019}. Overall the agreement with these observations is mixed with good agreement with the H$\beta$ + [O{\sc iii}]$\lambda$4959,5007 equivalent width distribution but with the observationally inferred line luminosity function offset to higher luminosities or space densities. One possible explanation to this discrepancy is that \citet{deBarros2019} measured higher FUV dust attenuation than predicted by \bluetides.

With the arrival of the \webbtelescope\ it will be possible to obtain direct measurements of individual line luminosities and equivalent widths for a large range of galaxies at $z\sim 8$ and beyond. Combined with other observational constraints this will allow us to test not only the base simulation but also the assumptions involved in modelling the nebular emission.

\subsection*{Acknowledgements}

We acknowledge support from NSF ACI-1036211, NSF AST-1009781, and NASA ATP grants NNX17AK56G and 80NSSC18K1015. The BlueTides simulation was run on facilities at the National Center for Supercomputing Applications. TDM acknowledges support from Shimmins and Lyle Fellowships at the University of Melbourne.

\bibliographystyle{mnras}
\bibliography{p} 

\appendix

\section{The Impact of Photoionisation Modelling Assumptions}\label{sec:photo}

In this section we describe, within the context of our nebular emission model, the impact of various assumptions on the hydrogen ionising (Lyman-continuum, LyC) photon production and subsequent line emission.

\subsection{The Production of Ionising Photons}\label{sec:photo.ion}

\begin{figure}
\centering
\includegraphics[width=20pc]{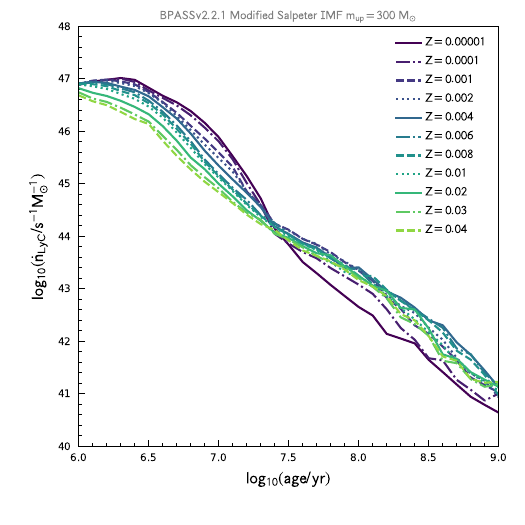}
\caption{The production rate of Lyman continuum (LyC) photons (H{\sc i} ionising) produced by a simple stellar population (SSP) per unit initial mass as a function of age for a range of different metallicities.}
\label{fig:LyC}
\end{figure}

\begin{figure}
\centering
\includegraphics[width=20pc]{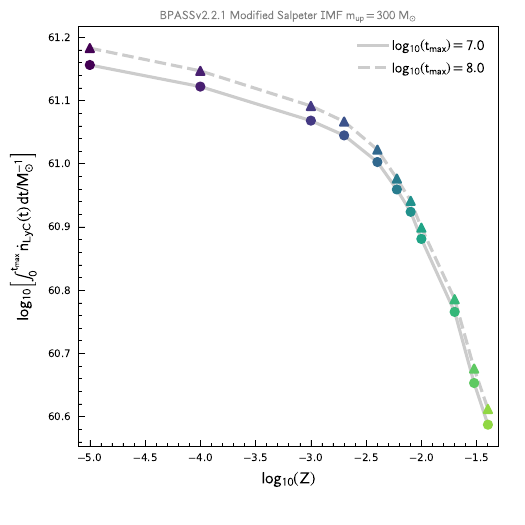}
\caption{The total number of LyC photons produced by SSP over the first 10 and 100 ${\rm Myr}$ as a function of metallicity.}
\label{fig:LyC_total}
\end{figure}

Young, massive, hot stars produce LyC photons. The photons can be reprocessed by surrounding gas (and dust) into nebular continuum and line emission. Consequently the production rate of these photons by a stellar population is sensitive to the joint distribution of stellar ages and metallicities. This is demonstrated in Fig. \ref{fig:LyC}, where we the LyC production rate $\dot{n}_{\rm LyC}$ as a function of age for a range of different metallicities assuming our default choices of SPS model and IMF is shown. The production rate drops rapidly after the first few million years at higher ages and metallicities, declining by $\sim 10-100$ as the population ages from $1\,\to 10\,{\rm Myr}$ and then again by a factor of $\sim 100$ from $t=10\,\to 100\,{\rm Myr}$. At young ages ($<20$ Myr) the lowest metallicity populations can produce up-to $10$ times as many LyC photons, though at later times this trend reverses. The overall difference in the production of LyC photons as a function of metallicity is summarised in Fig. \ref{fig:LyC_total} where we show the {\em total} number of LyC photons produced by an SSP from $t=0\to 10^{7}\,{\rm yr}$ and  $t=0\to 10^{8}\,{\rm yr}$. The lowest metallicity modelled ($Z=10^{-5}$) SSP considered produces approximately double the number LyC photons over its lifetime compared to $Z=0.01$.

\subsubsection{The Ionising Photon Hardness}\label{sec:photo.ion.hardness}

\begin{figure}
\centering
\includegraphics[width=20pc]{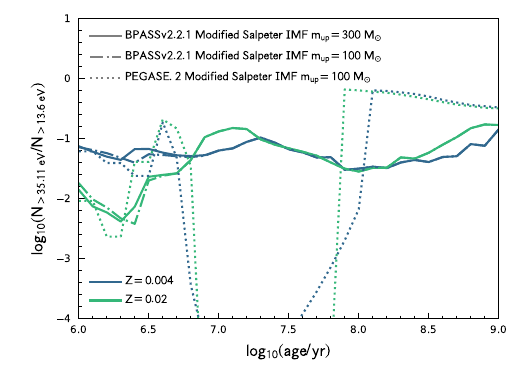}
\caption{The hardness of the ionising of photon spectrum (defined as the ratio of O{\sc ii} to H{\sc i} ionising photons) as a function of age for two metallicities and three SPS model / IMF combinations.}
\label{fig:hardness}
\end{figure}

More complex atoms have a range of potential ionisation states each excited by photons of different energies. For example, helium can be singly ionised by photons with $E_{\gamma}>24.6\,{\rm eV}$ and doubly ionised by those with $E_{\gamma}>54.4\,{\rm eV}$. For this reason it is useful to also consider the ionising photon hardness, essentially a ratio of the number of more energetic photons to $\dot{n}_{\rm LyC}$. The left panel of Fig. \ref{fig:hardness} shows the hardness the LyC by comparing the number of LyC and O{\sc ii} ionising ($>35.1\,{\rm eV}$) photons as a function of age for two metallicities $Z\in\{0.02, 0.004\}$. At the youngest ages the higher metallicity stellar population produces significantly fewer ($\sim 1$ dex) harder photons. For older ($>10\,{\rm Myr}$) populations the hardness is similar. The impact of this will be line ratios that vary as a function of the age of the ionising stellar population.

\subsubsection{The Effect of SPS Model Choice}\label{sec:photo.ion.SPS}

\begin{figure}
\centering
\includegraphics[width=20pc]{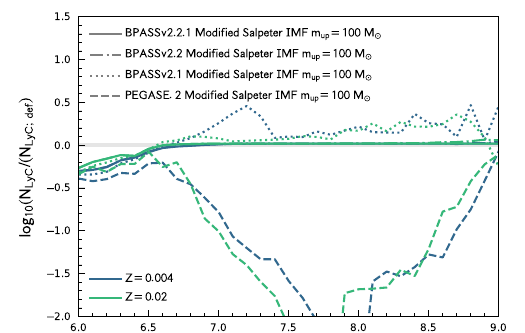}
\includegraphics[width=20pc]{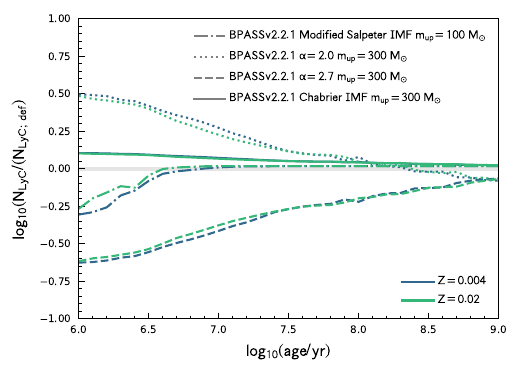}
\caption{The difference between the number of LyC photons produced assuming alternative SPS models (top) and IMFs (bottom) relative to our default modelling choices (BPASSv2.2.1, $\alpha_2=2.35$, $m_{\rm up}=300\,{\rm M_{\odot}}$).}
\label{fig:SPS_IMF}
\end{figure}

The number of LyC photons and the shape of the ionising spectrum predicted for a given stellar population is also sensitive to the a range of stellar evolution and atmosphere modelling assumptions and thus choice of stellar population synthesis (SPS) model \citep[see also][]{WLS2019}. The middle panel of Fig. \ref{fig:SPS_IMF} shows a comparison between $\dot{n}_{\rm LyC}$ for different SPS models/versions; these include the three most recent versions of {\sc bpass} (v2.2.1, v2.2, v2.1) and the {\sc Pegase.2} model \citep{Pegase}. This analysis reveals relatively small differences between the different {\sc bpass} versions but larger differences between {\sc bpass} and {\sc Pegase.2}. This difference is particularly acute at ages $>5$ Myr where the LyC production rate predicted by {\sc Pegase.2} drops of much more rapidly than in {\sc bpass}. The left panel of Fig. \ref{fig:SPS_IMF} shows the difference in the hardness between the default model and {\sc bpass}; again, the most notable feature is the difference at $>5$ Myr.

\subsubsection{The Effect of the Choice of IMF}\label{sec:photo.ion.IMF}

Both the production rate and hardness are also affected by the choice of IMF. The right hand panel of Fig. \ref{fig:SPS_IMF} shows the production rate relative to our default model for several different high-mass slopes $\alpha\in\{2.0, 2.35, 2.7\}$ and for a lower ($100\,{\rm M_{\odot}}$) high-mass cut-off. Assuming a shallower slope ($\alpha = 2.0$) yields more around double the number of LyC photons overall with the enhancement decreasing with age. Assuming a steeper slope has the opposite effect albeit with a slightly larger magnitude. Adopting a lower high-mass cut-off reduces the number of LyC photons produced at the youngest ages, overall leading to around $\sim 30-50\%$ less LyC photons produced by the SSP over its lifetime, depending on the metallicity.

\subsection{Photoionisation Modelling}\label{sec:photo.photo}

Using the modelling procedure described above we now make predictions for line luminosity and equivalent widths. We concentrate here on 12 prominent UV and optical lines. In making these predictions we assume a constant star formation history with fixed metallicity.

\begin{figure}
\centering
\includegraphics[width=20pc]{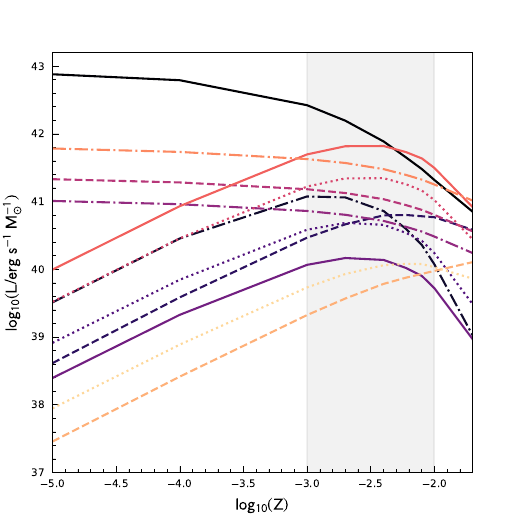}
\includegraphics[width=20pc]{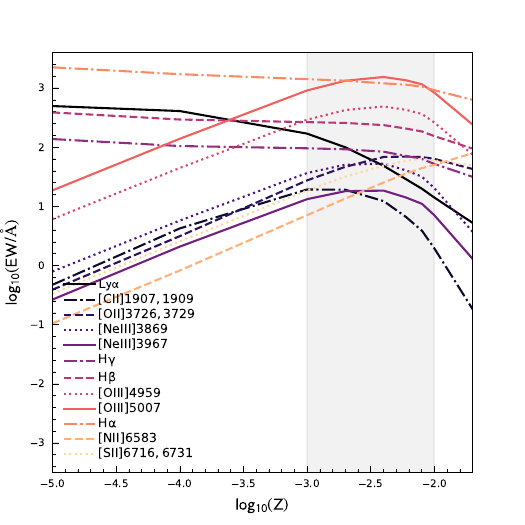}
\caption{The predicted intrinsic luminosity (top) and equivalent width (bottom) as a function of metallicity for a range of prominent emission lines in the rest-frame UV and optical. In both cases we assume constant star formation for 10 Myr. The thick grey band denotes the rough range of metallicities predicted by \bluetides\ for galaxies with $M^{*}>10^{8}\,{\rm M_{\odot}}$.}
\label{fig:lines_combined}
\end{figure}

Fig. \ref{fig:lines_combined} shows the predicted line luminosities (per unit stellar mass) and equivalent widths (EWs) for a range of prominent rest-frame UV and optical emission lines as a function of metallicity. In both cases we assume continuous star formation for 10 Myr. The luminosity of the hydrogen lines largely track the change in the LyC production rate with metallicity with the luminosity dropping by $\sim 0.5$ dex over the metallicity range considered. The non-hydrogen lines exhibit more complicated behaviour with an increase to $Z\sim 10^{-2.5}$ before declining to higher metallicities. The rapid increase broadly reflects the increasing abundance of each element in the ISM while the drop at high metallicities reflects the decline in the number of suitably energetic photons. The metallicity dependence of the EW of each line exhibits a similar behaviour, albeit often with reduced magnitude.

\begin{figure}
\centering
\includegraphics[width=20pc]{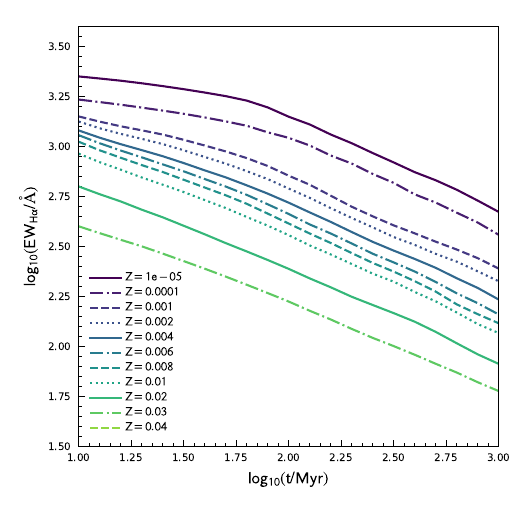}
\caption{The evolution of the H$\alpha$ equivalent width assuming continuous star formation for a range of metallicities.}
\label{fig:line_EWs_age_HI6563}
\end{figure}

The equivalent width of any line is also sensitive the star formation history of the stellar population. Fig. \ref{fig:line_EWs_age_HI6563} shows how the equivalent width of H$\alpha$ varies with the duration of continuous star formation for a range of metallicities. The equivalent width declines from $\sim 1000-2000{\rm\AA}$ after to 10 Myr to $\sim 150-300{\rm\AA}$ after 1 Gyr.

\subsubsection{The Effect of Photoionisation Modelling Assumptions}\label{sec:photo.photo.modelassumptions}

\begin{figure*}
\centering
\includegraphics[width=45pc]{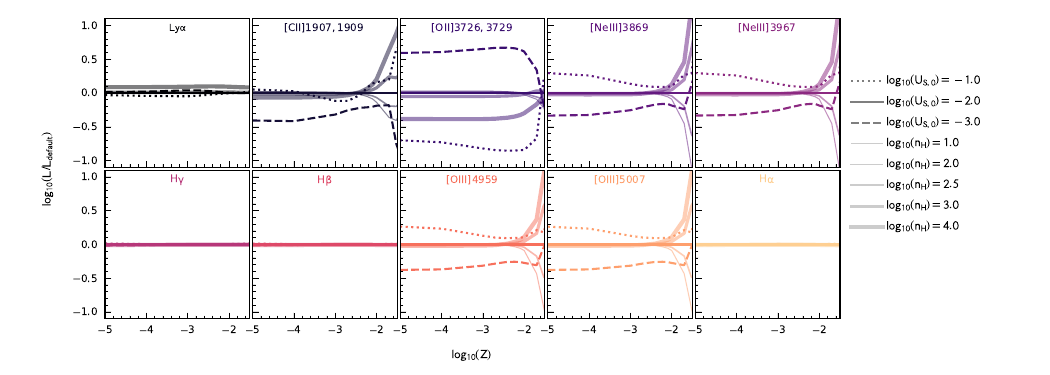}
\caption{The effect of changing the ionisation parameter $U_{S,{\rm ref}}$ and hydrogen density $n_{H}$ of the relative luminosity of each of the emission lines considered in this work. The default model assumes $\log_{10}(U_{S,{\rm ref}})=-2$ and $\log_{10}(n_{H}/cm^{-3})=2.5$.}
\label{fig:photo_params}
\end{figure*}

\begin{figure}
\centering
\includegraphics[width=20pc]{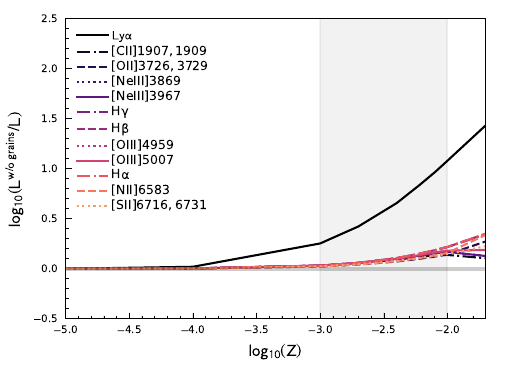}
\caption{The impact of turning off grain physics on the line luminosity assuming 10 Myr constant star formation.}
\label{fig:const_dust_comparison}
\end{figure}

In addition to the choice of SPS model and IMF the strength of lines are also senstive to the parameters encapsulating the geometry, density, and excitation of the H {\sc ii} region in addition to the presence of dust grains. Fig. \ref{fig:photo_params} demonstrates the effect on changing both the reference ionisation parameter $U_{S,{\rm ref}}$ and hydrogen density $n_{H}$ on the strengths of the same 12 prominent lines considered previously. While the hydrogen lines are largely insensitive to these choices many of the other lines, and in particular line ratios, are strongly sensitive with the effect dependent on the metallicity. The incusion of dust-grains in the H {\sc ii} region not only provide an additional source of LyC photon attenuation but also play a role in photoelectric heating of the gas. Fig. \ref{fig:const_dust_comparison} shows the predicted line luminosities when grains are omitted from the model. Omitting grains generally results in stronger lines with the effect being strongly metallicity dependent.

\section{Star Formation History Sampling Effects}\label{sec:appendix.sampling}

\begin{figure}
\centering
\includegraphics[width=20pc]{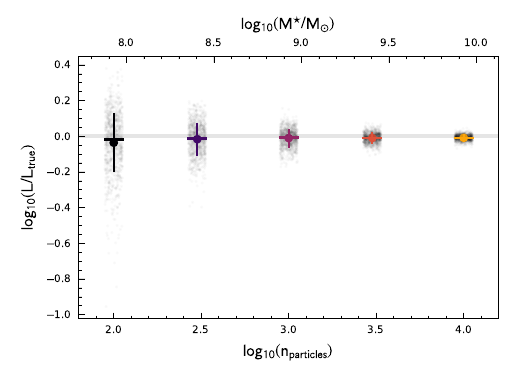}
\caption{The ratio of the modelled line luminosity compared to the true luminosity as a function of the number of particles used to sample the star formation history. The top-axis shows the corresponding stellar mass assuming the mean \bluetides\ stellar particle mass.}
\label{fig:sampletest}
\end{figure}

The Lyman continuum photon production rate is a strong function of the age, and to a lesser extent, metallicity of the stellar population. As the star formation history of each galaxy is sampled, at the lowest masses considered, by a small ($\sim 100$) number of individual star particles, this raises the possibility that the predicted line properties differs from the truth because of SFH sampling affects. To gauge the impact of this effect we re-sample the average star formation history of galaxies at $z=8$ using different numbers of fixed mass star particles ($n=10^2-10^4$); corresponding roughly to stellar masses of $10^8-10^{10}\,{\rm M_{\odot}}$. The result of this analysis is shown in Fig. \ref{fig:sampletest}. This test reveals that there is no significant bias in the average (median or mean) of the predicted line luminosity (in this case H$\alpha$), even at the lowest masses considered in this study. However, at low-masses there is some scatter ($\approx 0.1$ dex for $n=300$ particles / $M_{\star}\approx 2.5\times 10^{8}\,{\rm M_{\odot}}$). Because of the steepness of the luminosity function this will have the effect of flattening the LF at the lowest-luminosities.

\section{Detailed Predictions for Individual Lines}\label{sec:detailed_predictions}

\begin{figure*}
\centering
\includegraphics[width=40pc]{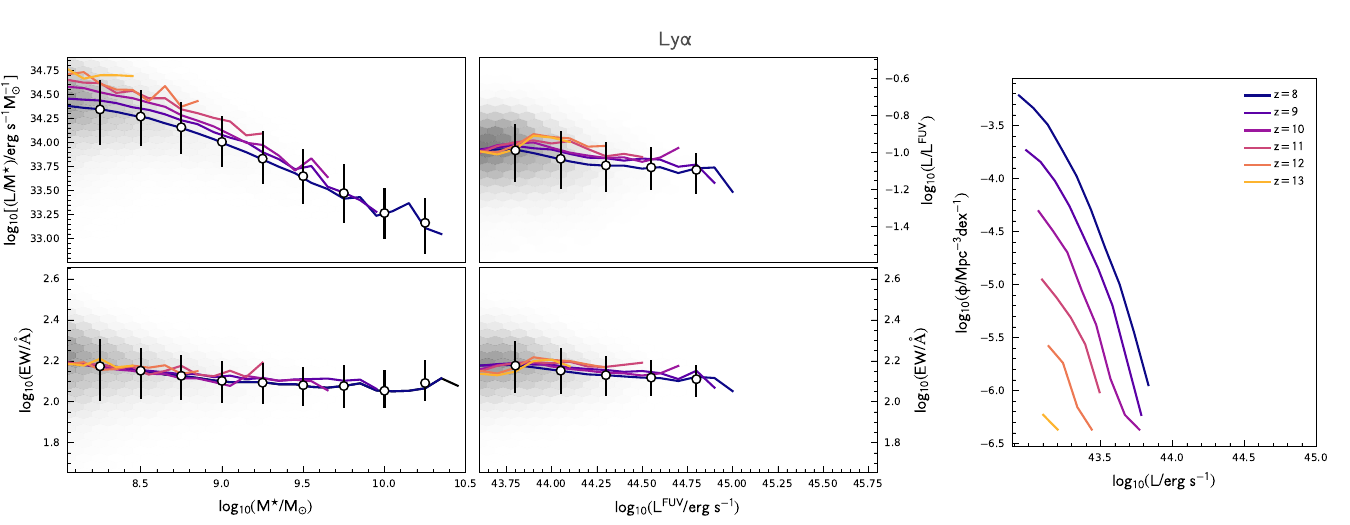}
\includegraphics[width=40pc]{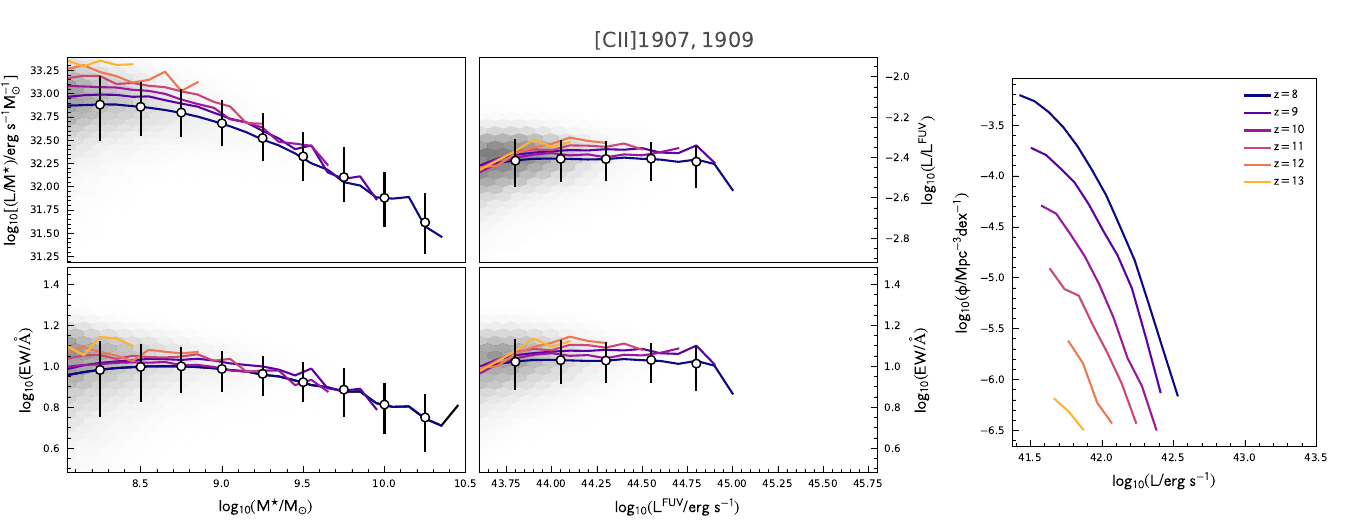}
\includegraphics[width=40pc]{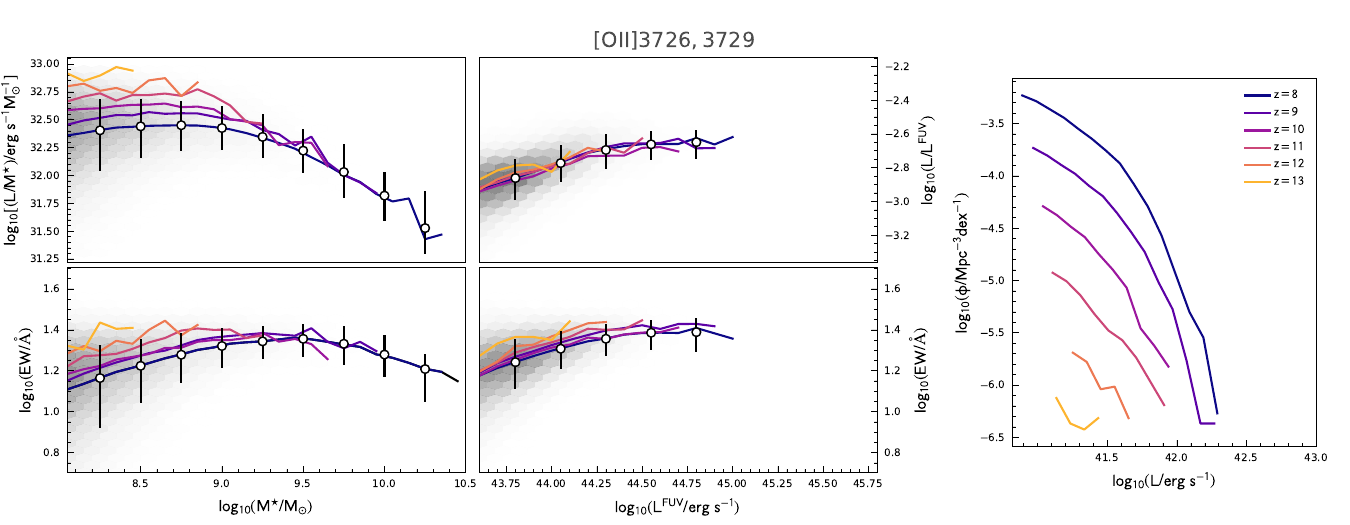}
\caption{The relationship between line luminosity and equivalent width, and stellar mass and far-UV luminosity and line luminosity function (right-hand panels) at $z=8-13$ for Lyman-$\alpha$, [C{\sc iii}],C{\sc iii}]$\lambda 1907,1090$, and [O{\sc ii}]$\lambda 2726,3729$.}
\label{fig:individual_set_1}
\end{figure*}

\begin{figure*}
\centering
\includegraphics[width=40pc]{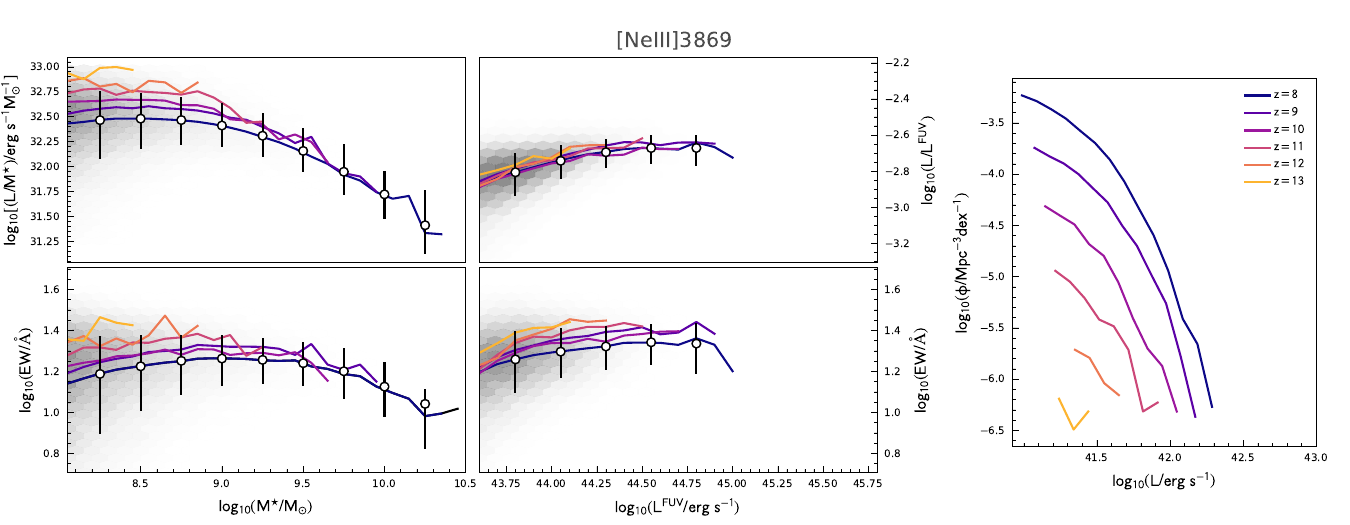}
\includegraphics[width=40pc]{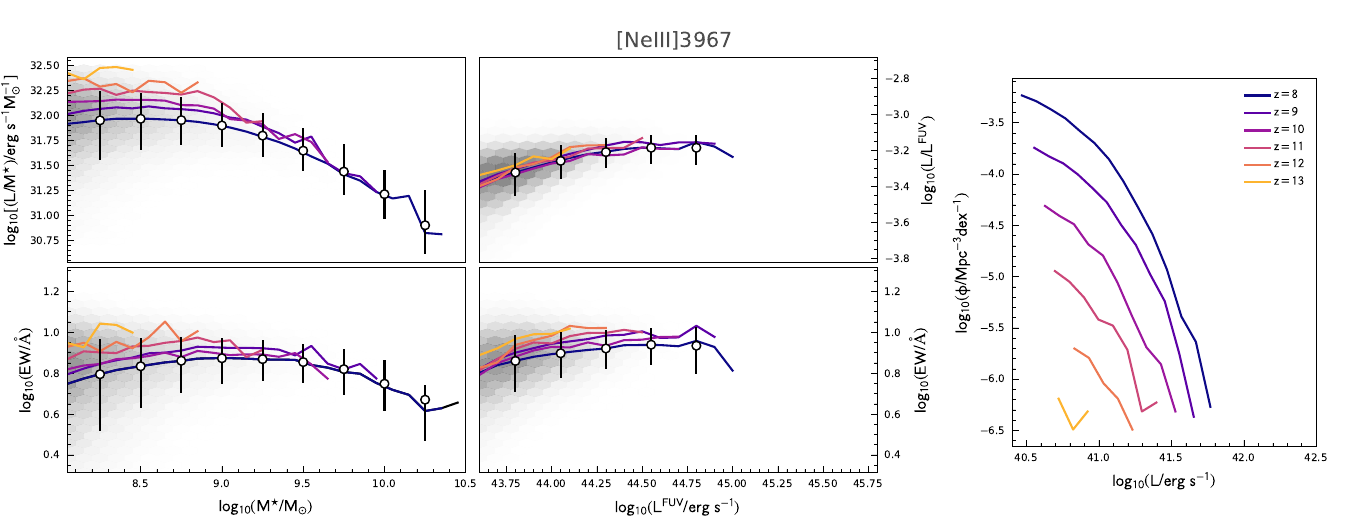}
\includegraphics[width=40pc]{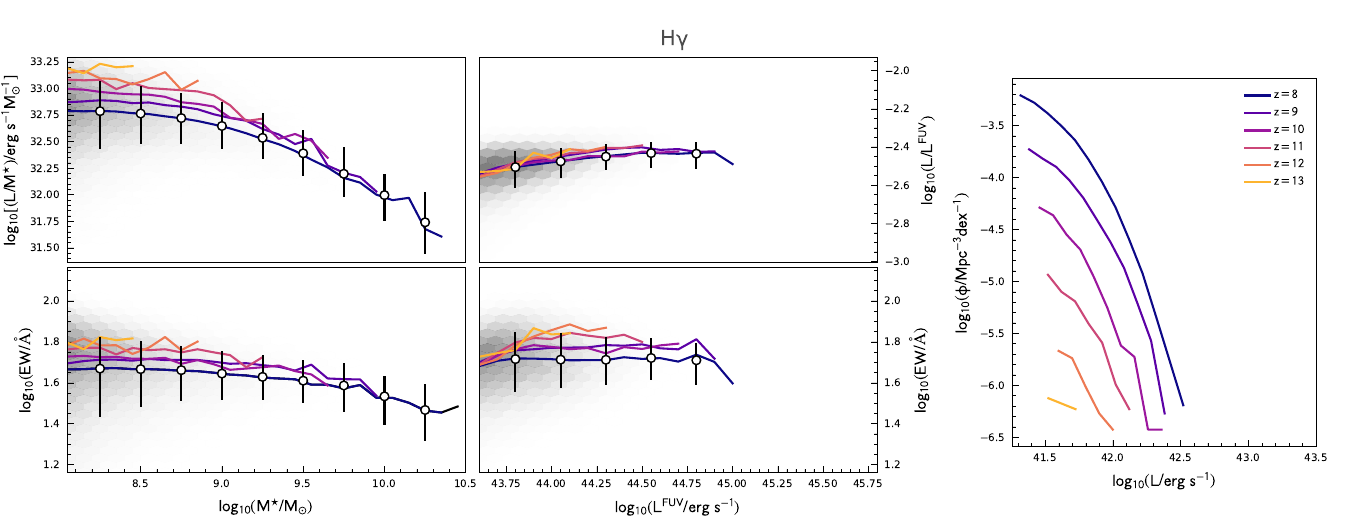}
\caption{The same as Fig. \ref{fig:individual_set_1} but for the [Ne{\sc iii}]$\lambda 3869$,  [Ne{\sc iii}]$\lambda 3967$, and H$\gamma$ lines.}
\label{fig:individual_set_2}
\end{figure*}

\begin{figure*}
\centering
\includegraphics[width=40pc]{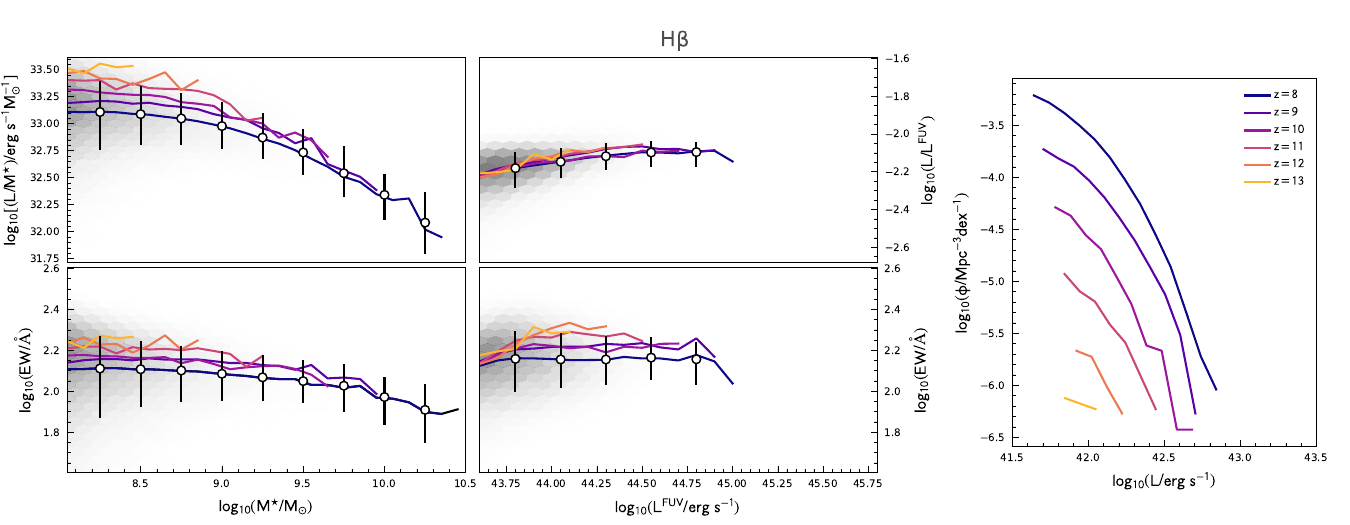}
\includegraphics[width=40pc]{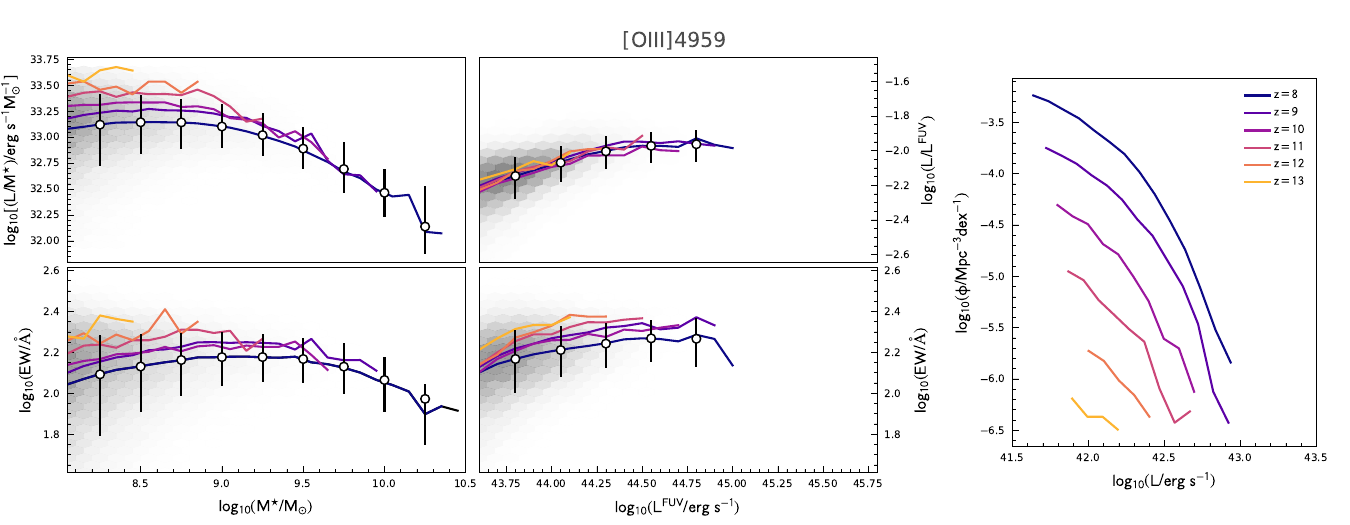}
\includegraphics[width=40pc]{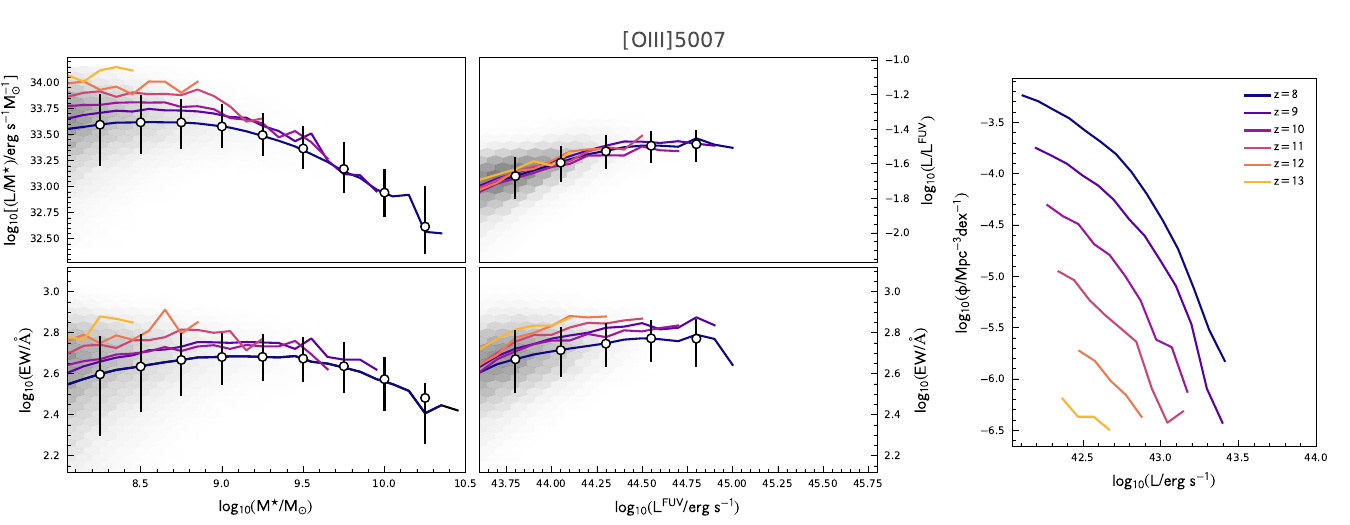}
\caption{The same as Fig. \ref{fig:individual_set_1} but for the H$\beta$,  [O{\sc iii}]$\lambda 4959$, and [O{\sc iii}]$\lambda 5007$ lines.}
\label{fig:individual_set_3}
\end{figure*}

\begin{figure*}
\centering
\includegraphics[width=40pc]{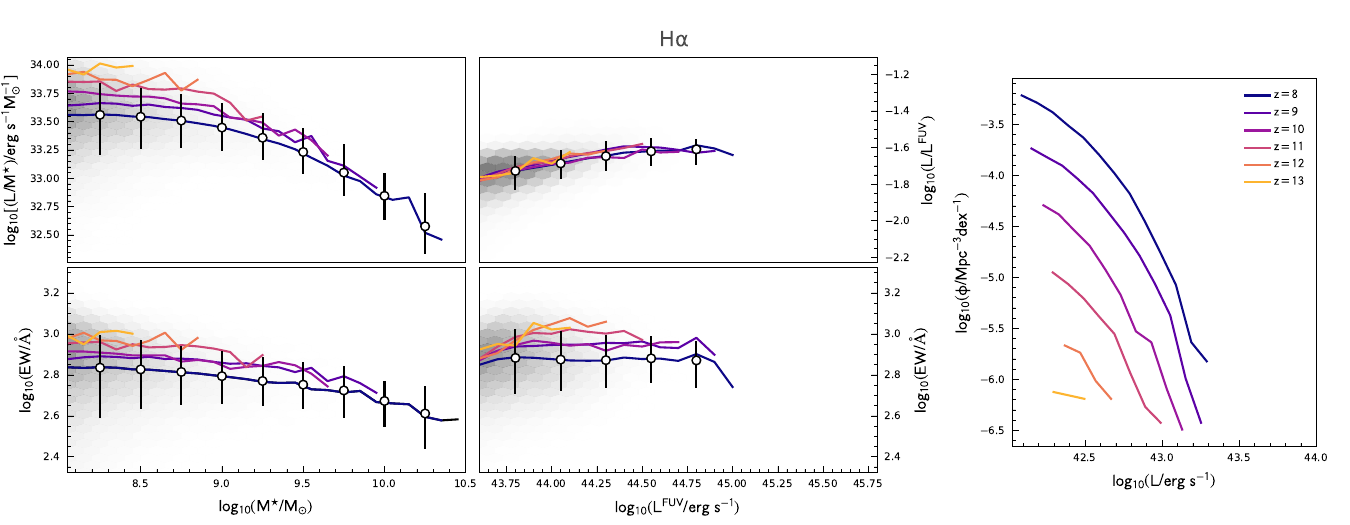}
\includegraphics[width=40pc]{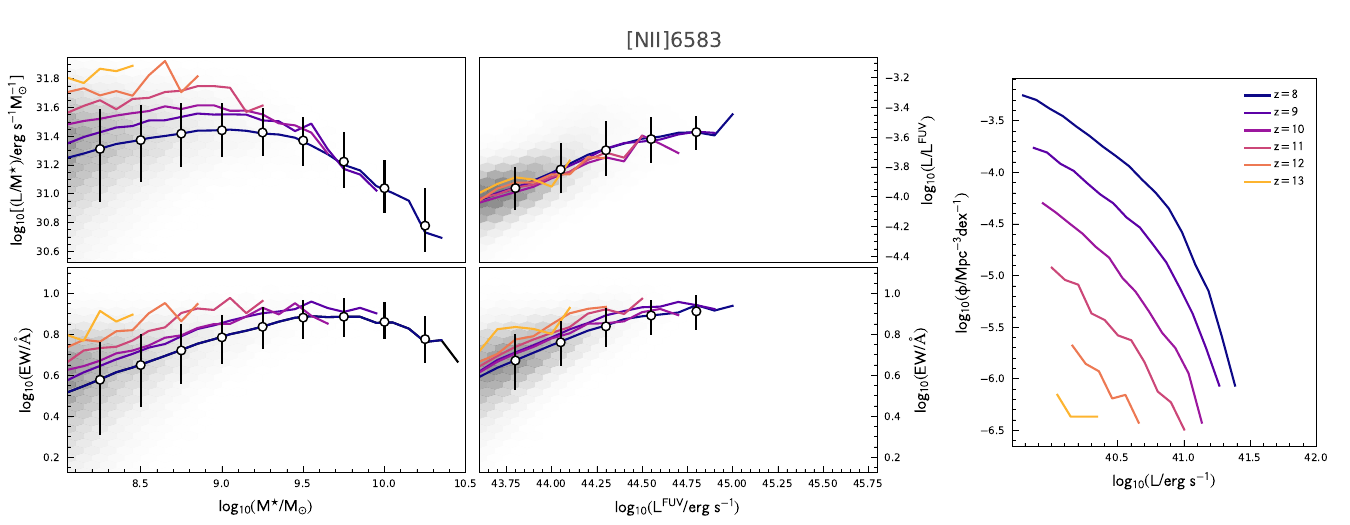}
\includegraphics[width=40pc]{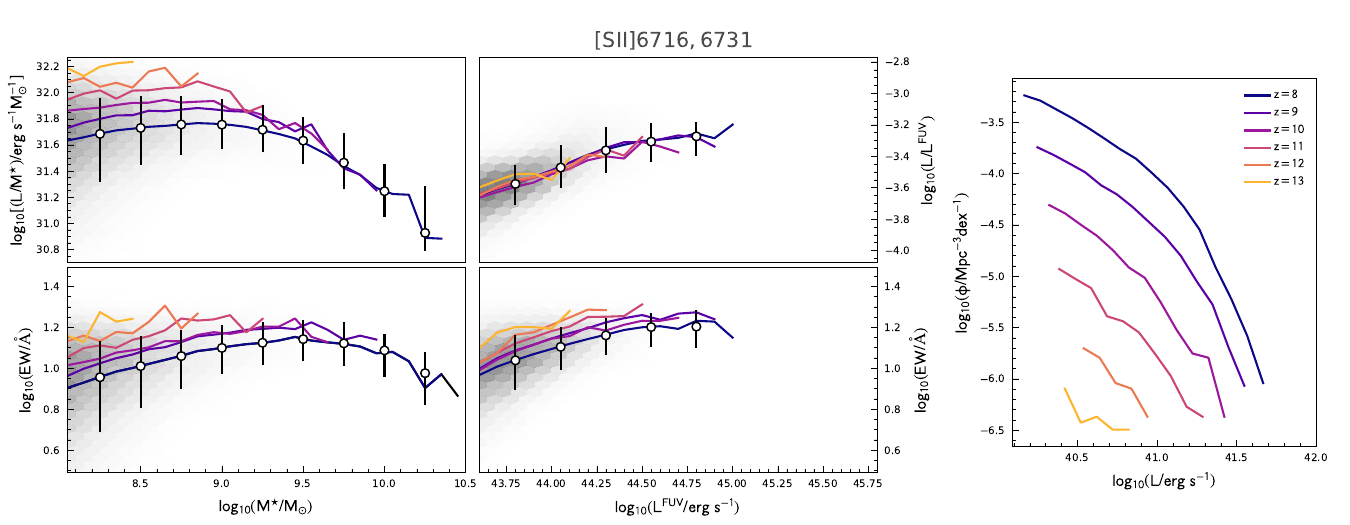}
\caption{The same as Fig. \ref{fig:individual_set_1} but for the H$\alpha$, N{\sc ii}$\lambda 6583$, and [S{\sc ii}]$\lambda 6716,6731$ lines.}
\label{fig:individual_set_4}
\end{figure*}

In Figs. \ref{fig:individual_set_1}-\ref{fig:individual_set_4} we show the EW and luminosity as a function of stellar mass and observed UV luminosity for each line in addition to the luminosity function. Tabulated values for each line and redshift are available at \url{https://github.com/stephenmwilkins/BluetidesEmissionLines_Public}.

\end{document}